\documentclass[5p,twocolumn]{elsarticle}
\journal{Physics Letters B}

\usepackage{amsmath,amssymb,mathrsfs,bm}
\usepackage{url}
\usepackage{hyperref}
\hypersetup{
colorlinks=true,
linkcolor=blue,
anchorcolor =red,
citecolor=blue,
filecolor = red,
urlcolor=blue,
pdfauthor=author}
\usepackage{txfonts}
\usepackage[utf8]{inputenc}
\usepackage{comment}
\def\av<#1>{\left\langle\,#1\,\right\rangle}
\def\ev<#1>{\left\langle\,#1\,\right\rangle_{\rm{ev}}}
\DeclareMathOperator{\diag}{diag}

\begin{document}

\begin{frontmatter}

\title{Hydrodynamic fluctuations and ultra-central flow puzzle \\ in heavy-ion collisions\tnoteref{reportnum}}
\tnotetext[reportnum]{Report number: YITP-23-54}

\author[Sophia]{Kenshi Kuroki\corref{cor}}
\ead{k-kuroki-e23@eagle.sophia.ac.jp}
\author[Hiroshima,Sophia]{Azumi Sakai}
\ead{azumi-sakai@hiroshima-u.ac.jp}
\author[Kyoto]{Koichi Murase}
\ead{koichi.murase@yukawa.kyoto-u.ac.jp}
\author[Sophia]{Tetsufumi Hirano}
\ead{hirano@sophia.ac.jp}
\cortext[cor]{Corresponding author}

\affiliation[Sophia]{organization={Department of Physics, Sophia University},
            city={Tokyo},
            postcode={102-8554},
            country={Japan}}
\affiliation[Hiroshima]{organization={Physics Program, Graduate School of Advanced Science and Engineering, Hiroshima University},
            city={Hiroshima},
            postcode={739-8526},
            country={Japan}}
\affiliation[Kyoto]{organization={Yukawa Institute for Theoretical Physics, Kyoto University},
            city={Kyoto},
            postcode={606-8502},
            country={Japan}}

\begin{abstract}
One of the long-standing problems 
in the field of high-energy heavy-ion collisions is that the dynamical models based on viscous hydrodynamics fail to describe the experimental elliptic flow $v_2$ and the triangular flow $v_3$ simultaneously in ultra-central collisions.
The problem, known as the \textit{ultra-central flow puzzle}, is specifically that hydrodynamics-based models predict the flow ratio of the two-particle cumulant method $v_2\{2\}/v_3\{2\} > 1$ while $v_2\{2\}/v_3\{2\} \sim 1$ in the experimental data.
In this Letter, we focus on the effects of hydrodynamic fluctuations during the space-time evolution of the QGP fluid on the flow observables in the ultra-central collisions.
Using the (3+1)-dimensional integrated dynamical model which includes relativistic fluctuating hydrodynamics, we analyze the anisotropic flow coefficients $v_n\{2\}$ in 0--0.2\% central Pb+Pb collisions at $\sqrt{s_\text{NN}}=2.76~\text{TeV}$\@.
We find that the hydrodynamic fluctuations decrease the model overestimate of $v_2\{2\}/v_3\{2\}$ from the experimental data by about 19\% within the present setup of $\eta/s = 1/2\pi$.
This means that the hydrodynamic fluctuations qualitatively have an effect to improve the situation for the puzzle, but the effect of the hydrodynamic fluctuations alone is quantitatively insufficient to resolve the puzzle.
The decrease of the ratio largely depends on the shear viscosity $\eta/s$, which calls for future comprehensive analyses with, for example, a realistic temperature-dependent viscosity.
\end{abstract}


\end{frontmatter}

Various observations indicate that the dynamics of the quark--gluon plasma (QGP) created in high-energy heavy-ion collisions at the Relativistic Heavy Ion Collider (RHIC) and the Large Hadron Collider (LHC) are well described by relativistic hydrodynamics (see e.g. Ref.~\cite{Jaiswal:2016hex} for a review).
The dynamical models based on relativistic hydrodynamics have been successful in describing the anisotropic flow coefficients $v_n$~\cite{Qin:2010pf, Schenke:2011tv, Qiu:2011iv, Shen:2011eg, Bozek:2011ua, Gardim:2012yp, Hirano:2012kj}.
Recently, the transport properties of the QGP, which are difficult to calculate from the first principles of quantum chromodynamics (QCD)~\cite{Meyer:2011gj, Bazavov:2019lgz, Ghiglieri:2018dib}, have been systematically constrained by means of Bayesian inference using dynamical models based on viscous hydrodynamics~\cite{Bernhard:2016tnd, Bernhard:2019bmu, Moreland:2018gsh, JETSCAPE:2020shq, JETSCAPE:2020mzn, Nijs:2020roc, Auvinen:2020mpc, Parkkila:2021tqq}.
In particular, $v_n$ are the crucial observables to constrain the shear and bulk viscosity of QGP.

Besides, there is a problem called the  ``ultra-central flow puzzle'' that no hydrodynamics-based models have ever reproduced the data of the elliptic flow $v_2$ and the triangular flow $v_3$ at the same time in the ultra-central collisions~\cite{Giannini:2022bkn}.
The flow coefficients $v_2\{2\}$ and $v_3\{2\}$ from the two-particle cumulant method were reported to be almost the same in ultra-central collisions at LHC~\cite{ALICE:2011ab, ATLAS:2012at, CMS:2013bza, ALICE:2018rtz, ATLAS:2019peb}.
However, existing dynamical models predict larger $v_2\{2\}$ than $v_3\{2\}$ in the ultra-central collisions failing to reproduce the experimental behavior $v_2\{2\} \sim v_3\{2\}$.
This contradicts our expectation that models based on hydrodynamics perform better in central collisions because of larger multiplicities and volumes of locally thermalized domains.
This implies that the state-of-the-art dynamical models, which are also intensively used in Bayesian analyses, potentially have missing pieces.
So far, a number of attempts have been made to resolve the puzzle from different aspects, such as improved descriptions of initial conditions~\cite{Luzum:2012wu, Denicol:2014ywa, Shen:2015qta, Bhalerao:2015iya, Loizides:2016djv, Gelis:2019vzt, Carzon:2020xwp, Snyder:2020rdy, Zakharov:2020irp}, effects of the transport coefficients~\cite{Luzum:2012wu, Rose:2014fba, Shen:2015qta, Plumari:2015cfa}, and the equations of state~\cite{Alba:2017hhe}, but none has yet succeeded in explaining the experimental behavior $v_2\{2\} \sim v_3\{2\}$ within the dynamical models based on hydrodynamics.

In ultra-central collisions, an approximate linear mapping from the initial geometrical anisotropies $\varepsilon_n$ to the anisotropic flow coefficient $v_n$ is known~\cite{Noronha-Hostler:2015dbi, Sievert:2019zjr, Rao:2019vgy},
\begin{equation}
    v_n = \kappa_n \varepsilon_n.
\end{equation}
Although initial anisotropies of the second and the third orders are almost the same $\varepsilon_2 \sim \varepsilon_3$ in the initial stage, the viscosity leads to the ordering $\kappa_2 > \kappa_3$, and thus $v_2 > v_3$.
This is because smaller structures, which are associated with higher anisotropies, are smeared out more by the viscosity~\cite{Carzon:2020xwp, Snyder:2020rdy}.
However, $\varepsilon_n$ and $v_n$ can be de-correlated by additional event-by-event fluctuations in the hydrodynamic and hadronic stages.
This happens more easily in ultra-central collisions due to the smaller geometrical origin of the initial anisotropies.
The effect of the initial fluctuations has been intensively investigated in existing studies, but the effects of dynamical fluctuations arising in the hydrodynamic and later stages have not been comprehensively studied so far.

In this Letter, we focus on hydrodynamic fluctuations~\cite{Landau1959fluid, lifshitz1980statistical} arising in the hydrodynamic stage.
Hydrodynamic fluctuations are thermal fluctuations related to the hydrodynamic description of the system.
In hydrodynamics, microscopic degrees of freedom are integrated out by coarse-graining so that the system is described by only a few slow macroscopic variables.
However, macroscopic dynamics cannot be completely separated from microscopic one.
The microscopic dynamics induces the thermal fluctuations of the macroscopic variables on an event-by-event basis, which is nothing but the hydrodynamic fluctuations.
Since the hydrodynamic fluctuations and the dissipations are mutually related by the fluctuation--dissipation relation (FDR)~\cite{Kubo:1957mj}, it is natural and indispensable to consider hydrodynamic fluctuations in the hydrodynamic models for the highly non-equilibrium dynamics of the heavy-ion collisions~\cite{Kapusta:2011gt}.
Previous studies have shown that causal hydrodynamic fluctuations~\cite{Murase:2013tma, Young:2013fka, Murase:2015oie, Murase:2016rhl, Murase:2019cwc} play an important role to explain the centrality dependence of the rapidity decorrelation of anisotropic flows at LHC~\cite{Sakai:2020pjw, Sakai:2021rug}.
Since hydrodynamic fluctuations disturb fluid evolution randomly, they are expected to increase fluctuations of $v_n$ and affect $v_n\{2\}$, especially in ultra-central collisions.
Also, hydrodynamic fluctuations are more likely to affect higher-order $v_n\{2\}$, which are related to smaller structures~\cite{Murase:2016rhl, Singh:2018dpk}, and thus could be a key to resolving the puzzle.

We use the natural units $\hbar=c=k_B=1$ and the sign convention of the metric $g_{\mu\nu}=\diag(+, -, -, -)$ throughout this Letter.

\vspace{10pt}
We employ the (3+1)-dimensional integrated dynamical model~\cite{Murase:2016rhl, Sakai:2020pjw} with hydrodynamic fluctuations to describe the space-time evolution of high-energy heavy-ion collisions.
In the integrated dynamical model, as prescribed in Ref.~\cite{Hirano:2012kj}, we combine three models corresponding to the three stages of the collision reactions:
a Monte-Carlo version of the Glauber model (MC-Glauber)~\cite{Glauber:2006gd} smoothly extended in the longitudinal direction~\cite{Hirano:2005xf} for the initial entropy deposition, a relativistic fluctuating hydrodynamic model \texttt{rfh}~\cite{Murase:2015oie} which is relativistic dissipative hydrodynamics with hydrodynamic fluctuations for the space-time evolution of locally thermalized matters, and a hadron cascade model \texttt{JAM}~\cite{Nara:1999dz} for the microscopic transport of hadron gases.

The main dynamical equations of the causal second-order fluctuating hydrodynamics are the conservation law of the energy-momentum tensor $T^{\mu\nu}$ of fluids and the constitutive equation for dissipative currents.
We neglect the conservation law of the baryon number since we focus on heavy-ion collisions at LHC energies, at which the net baryon number is almost negligible around midrapidity.
The energy-momentum tensor $T^{\mu\nu}$ can be tensor-decomposed as follows:
\begin{equation}
    \label{eq:conservation}
    T^{\mu\nu} = eu^\mu u^\nu - P \Delta^{\mu\nu} + \pi^{\mu\nu},
\end{equation}
where $e$, $P$, and $\pi^{\mu\nu}$ are the energy density, the pressure, and the shear-stress tensor, respectively.
The flow velocity $u^\mu$ in the Landau frame is defined as $T^\mu{}_{\nu} u^\nu = e u^\mu$, and $\Delta^{\mu\nu} \coloneqq g^{\mu\nu} - u^\mu u^\nu$ is a projector for four-vectors onto the components transverse to $u^\mu$.
We do not consider the bulk pressure in the present study.
For an equation of state, we adopt \texttt{$s95p$-v1.1}~\cite{Huovinen:2009yb} which smoothly combines the equation of state from (2+1)-flavor lattice QCD simulations with that from the hadron resonance gas model corresponding to the hadron cascade model \texttt{JAM}.
The constitutive equations for $\pi^{\mu\nu}$ in this study are~\cite{Murase:2015oie, Murase:2019cwc, Baier:2007ix}
\begin{multline}
    \label{eq:constitutive}
    \tau_\pi \Delta^{\mu\nu}{}_{\alpha\beta} u^\lambda \partial_\lambda \pi^{\alpha\beta}
    + \left( 1+ \frac{4}{3} \tau_\pi \partial_\lambda u^\lambda \right) \pi^{\mu\nu} \\
    = 2\eta \Delta^{\mu\nu}{}_{\alpha\beta} \partial^\alpha u^\beta + \xi^{\mu\nu}_\pi,
\end{multline}
where $\eta$ and $\tau_\pi$ are the shear viscosity and the relaxation time, respectively.
The tensor $\Delta^{\mu\nu}{}_{\alpha\beta} \coloneqq \frac{1}{2} \left( \Delta^\mu{}_{\alpha} \Delta^\nu{}_{\beta} + \Delta^\nu{}_{\alpha} \Delta^\mu{}_{\beta} \right) - \frac{1}{3} \Delta^{\mu\nu} \Delta_{\alpha\beta}$ is a projector for second-rank tensors onto the symmetric and traceless components transverse to $u^\mu$.
The stochastic term $\xi^{\mu\nu}_\pi$ represents hydrodynamic fluctuations, whose magnitude is determined by the FDR.
In the  Milne coordinates $\left( \tau, \eta_{\mathrm{s}}, \bm{x}_\perp \right) \coloneqq \left( \sqrt{t^2-z^2}, {\rm tanh}^{-1}(z/t), x, y \right)$, the FDR is written as
\begin{multline}
    \label{eq:FDR}
    \langle \xi^{\mu\nu}_\pi\left( \tau, \eta_{\mathrm{s}}, \bm{x}_\perp \right) \xi^{\alpha\beta}_\pi\left( \tau', \eta'_{\mathrm{s}}, \bm{x}'_\perp \right) \rangle \\
    = 4 \eta T \Delta^{\mu\nu\alpha\beta} \frac{1}{\tau} \delta(\tau-\tau') \delta(\eta_{\mathrm{s}}-\eta'_{\mathrm{s}}) \delta^{(2)}(\bm{x}_\perp - \bm{x}'_\perp), 
\end{multline}
where $T$ is the temperature, and $\langle\cdots\rangle$ means the event average.
In the actual calculations, a spatial regularization is needed to tame the ultra-violet divergences caused in the non-linear hydrodynamic equations~\cite{Murase:2015oie}.
We employ the smearing of the noise fields by the Gaussian kernel of the widths $\lambda_\perp$ and $\lambda_{\eta_{\mathrm{s}}}$ in the transverse and longitudinal directions, respectively.
Smaller spatial cutoff parameters result in larger effects of hydrodynamic fluctuations.

For the initial conditions of hydrodynamic simulations, we generate the event-by-event entropy density distributions $s\left(\eta_{\mathrm{s}}, \bm{x}_\perp; \tau_0 \right)$ at the fixed hydrodynamic initial proper time $\tau_0$ using \texttt{mckln}~\cite{Hirano:2012kj}.
We parametrize the initial transverse profile from the linear combination of the participant number density of nuclei A and B, $\rho_{\rm part}^{\rm A}\left(\bm{x}_\perp\right)$ and $\rho_{\rm part}^{\rm B}\left(\bm{x}_\perp\right)$, respectively, and the number density of the binary collisions, $\rho_{\rm coll}\left(\bm{x}_\perp\right)$, for a randomly sampled impact parameter $b$ satisfying $P(b) db\propto b db$, as follows,
\begin{multline}
    \label{eq:Glauber}
    s \left(\bm{x}_\perp;  \tau_0 \right) \\
    = \frac{C}{\tau_0} \left\{ \frac{1-\alpha}{2} \left[\rho_{\rm part}^{\rm A}\left(\bm{x}_\perp\right) + \rho_{\rm part}^{\rm B}\left(\bm{x}_\perp\right)\right] + \alpha\rho_{\rm coll}\left(\bm{x}_\perp\right) \right\},
\end{multline}
where $C$ and $\alpha$ are the normalization factor and the hard fraction, respectively.
The initial transverse profiles are extended to the longitudinal direction using the modified Brodsky--Gunion--K\"uhn (BGK) model based on the idea of the \textit{rapidity trapezoid}~\cite{Hirano:2005xf, Brodsky:1977de, Adil:2005qn}.
In the present study, we neglect the initial transverse flow, the initial shear-stress tensor, and the initial longitudinal fluctuations.

At a switching temperature $T_\text{sw}$, we switch the description from the hydrodynamics to the microscopic kinetic theory using the Cooper--Frye formula~\cite{Cooper:1974mv} with a viscous correction~\cite{Teaney:2003kp, Monnai:2009ad}.
The subsequent space-time evolution of hadron gases including hadronic rescatterings and decays of resonances is described by the hadron cascade model \texttt{JAM}~\cite{Nara:1999dz}.

Let us summarize the parameters of the present study.
As in the previous calculations~\cite{Murase:2016rhl, Hirano:2012kj}, we set the initial proper time $\tau_0 = 0.6~\text{fm}$ and the switching temperature $T_\text{sw} = 155~\text{MeV}$.
For the transport properties of QGP, we choose the specific shear viscosity $\eta/s = 1/4\pi$~\cite{Kovtun:2004de} or $1/2\pi$ and the relaxation time $\tau_\pi = 3\eta/sT$~\cite{Song:2009gc, Baier:2007ix}.
We use the same initial parameters $C/\tau_0$ and $\alpha$ as the previous study~\cite{Sakai:2020pjw} for each hydrodynamic model to reproduce the centrality dependence of the experimental charged-particle multiplicity measured by the ALICE Collaboration~\cite{ALICE:2010mlf}.
These parameters for each hydrodynamic model are summarized in Table~\ref{tab:param}.
\begin{table}[tbp]
    \centering
    \caption{Parameters in hydrodynamic models.}
    \label{tab:param}
    \begin{tabular}{ c c c c c c } \hline
        Model         & $\eta/s$ & $\lambda_\perp~\text{(fm)}$ & $\lambda_{\eta_s}$ & $C/\tau_0$ & $\alpha$ \\ \hline
        Viscous hydro & $1/4\pi$ & N/A                         & N/A                & 49         & 0.13     \\
                      & $1/2\pi$ & N/A                         & N/A                & 47         & 0.13     \\ \hline
        Fluct. hydro  & $1/4\pi$ & 1.5                         & 1.5                & 41         & 0.16     \\
                      & $1/2\pi$ & 1.5                         & 1.5                & 36         & 0.16     \\ \hline\
    \end{tabular}
\end{table}

\vspace{10pt}
Because of the large computational cost of the event-by-event dynamical simulations including the (3+1)-dimensional hydrodynamics and hadronic cascades, it is impractical to perform the minimum-biased simulations and afterward select the 0--0.2\% events.
The simplest approach to effectively generate the ultra-central events would be to fix the impact parameter $b$ to 0~fm, but collisions with exact $b = 0~\text{fm}$ do not occur in reality.
Another approach would be to carry out the centrality selection at the initial stage using, e.g., the total entropy of the initial condition.
However, even with a fixed initial entropy, the measured centrality can fluctuate due to the non-trivial evolution in the hydrodynamic and hadronic stages, the finite acceptance, etc~\cite{Zhou:2018fxx}.
In this study, we employ the importance sampling method to reduce the overall simulation cost while determining the centrality in the final multiplicity as in experiments.
We introduce a weight function $w(b)=\exp\left(-b\right)$ into the impact parameter distribution as $P(b) db\propto b \cdot w(b) db$ and generate initial conditions with small $b$ intensively.
To cancel the artifact of the extra weight in the distribution so that the redistributed events reproduce the proper multiplicity distribution, we need reweighting in the statistical analyses as
\begin{equation}
    \label{eq:reweight}
    \ev<X> = \frac{\sum_i X_i / w(b_i)}{\sum_i 1/w(b_i)},
\end{equation}
where $\ev<\cdots>$ is an event average, and $X_i$ is a physical quantity in the event $i$ with impact parameter $b_i$.
Assuming that events with $b$ larger than 5 fm do not contribute to centrality 0--0.2\%, we generate only the initial conditions with $b$ smaller than 5 fm.
\begin{figure}[tbp]
    \centering
    \includegraphics[clip,width=\linewidth]{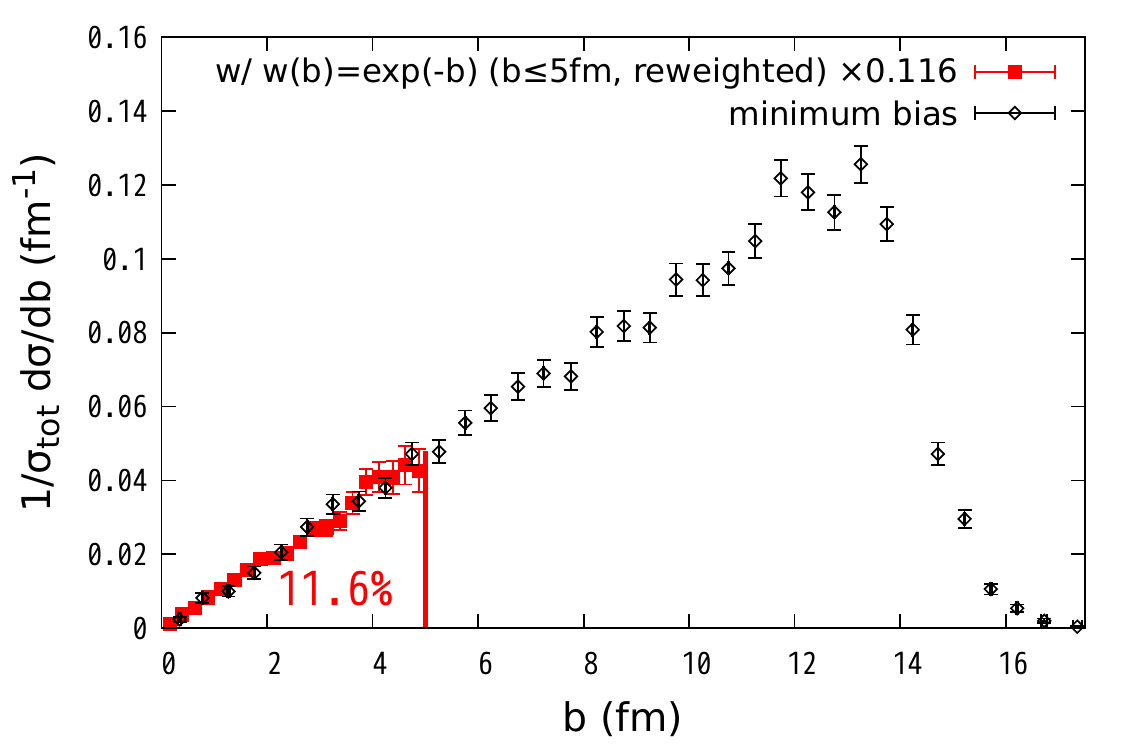}
    \caption{(Color Online)
            Impact parameter distribution in Pb+Pb collisions at $\sqrt{s_{\rm NN}} = 2.76~\text{TeV}$ based on the MC-Glauber model.
            Reweighted distribution with $b\leq5~\text{fm}$ (filled square) compared with minimum bias distribution (open diamond).
            }
    \label{fig:b_hist}
\end{figure}
Figure~\ref{fig:b_hist} shows that the cross-section with $b$ smaller than 5 fm is about 11.6\% of the total cross-section within our version of the MC-Glauber model.
Therefore, for the 0--0.2\% centrality events, we can select the events that belong to the top 1.72\% (= 0.2/11.6) of the reweighted multiplicity distribution.
The resulting fraction of the 0--0.2\% centrality events in the simulated number of events was about 10\%. This means that the total number of simulated events to get the same number of simulated 0--0.2\% events is reduced to about $(1/10\%)/(1/0.2\%) = 1/50$ compared to the case of minimum-biased simulations\footnote{
  Nevertheless, the actual computational cost would not reduce as much as 1/50 because the computational cost for the non-central collisions is usually much smaller than the central collisions.
  Also, even with the same number of simulated events, the final statistical error would be affected by the reweighting.}.
The impact parameter distribution of the centrality 0--0.2\% events has a peak at $\sim$0.8~fm. The maximum impact parameter within about 500 simulated events of centrality 0--0.2\% was less than 2.5 fm, which implies that the events of $b>2.5~\text{fm}$ essentially do not contribute to the 0--0.2\% centrality class.
This confirms that our choice of the initial cut of $b=0$--$5~\text{fm}$ is sufficiently large, and there is room to further reduce the initial cut below $5~\text{fm}$.

\vspace{10pt}
Using the integrated dynamical model with the above importance sampling, we perform simulations of Pb+Pb collisions at $\sqrt{s_{\mathrm{NN}}}=2.76~\text{TeV}$.
For each parameter set, we generate 5~000 hydrodynamic events and perform 10 independent particlizations and hadronic cascades for each hydrodynamic event, which ends up with 50~000 events in total.
We apply the multiplicity cut and obtain about 5~000 events as the 0--0.2\% central events.

The anisotropic flow coefficients from the two-particle cumulant method are calculated as
\begin{equation}
    \label{eq:vn}
    v_n\{2\}^2 = \ev<{\av<\mathrm{e}^{i n \Delta\phi}>}>
    ,
\end{equation}
where $\Delta\phi$ represents the difference of the azimuthal angles between two charged hadrons, and the inner $\av<\cdots>$ represents the average over the particle pairs in each event.
To compare the result with the data of the CMS Collaboration~\cite{CMS:2013bza}, we pick charged particles in the pseudo-rapidity range $|\eta_p|<2.4$ and the transverse momentum range $0.3<p_{\mathrm{T}}<3.0~\text{GeV}$ and calculate flows introducing a pseudo-rapidity gap $|\Delta\eta_p|_{\min}$.

\begin{figure}[tbp]
    \centering
    \includegraphics[clip,width=0.49\linewidth]{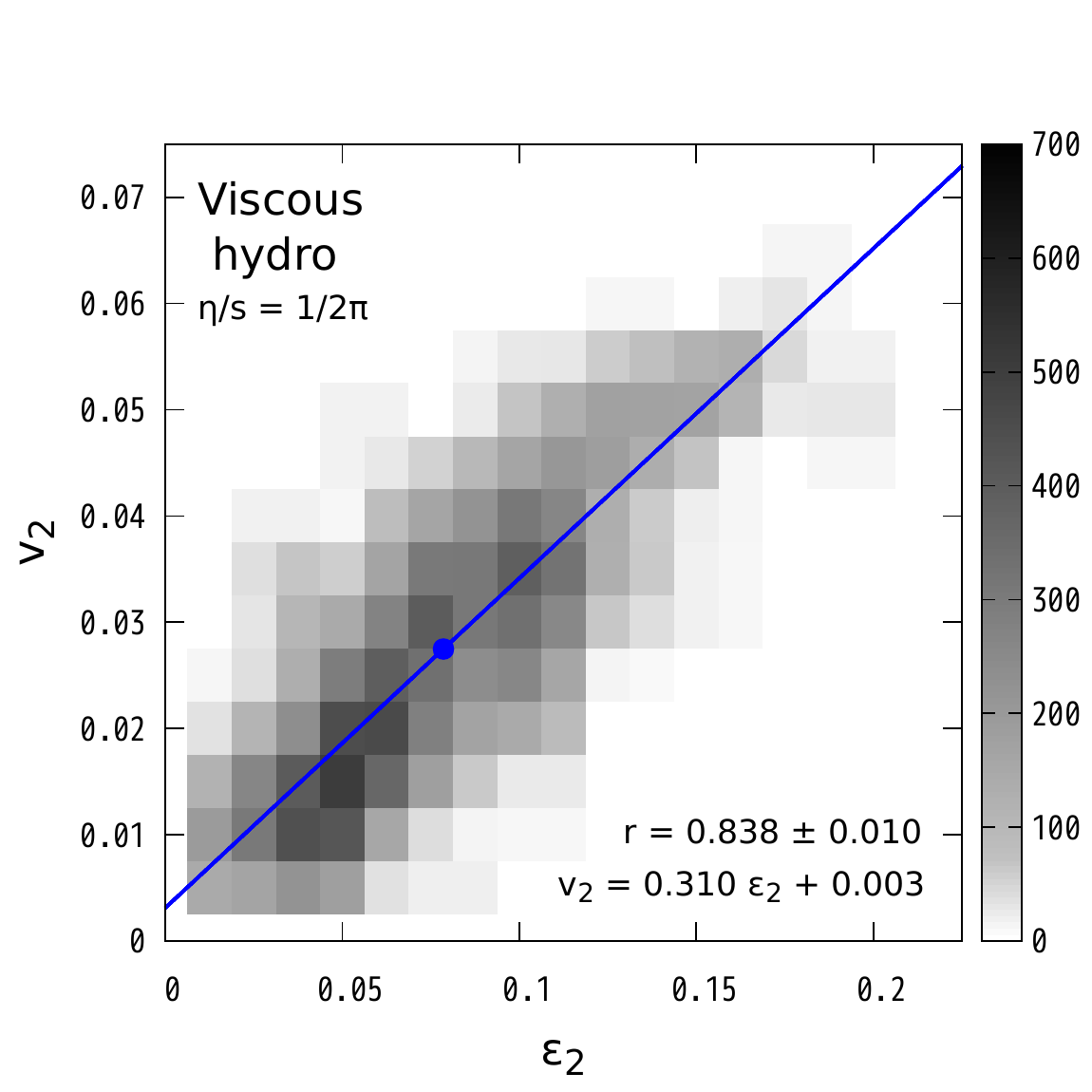}
    \includegraphics[clip,width=0.49\linewidth]{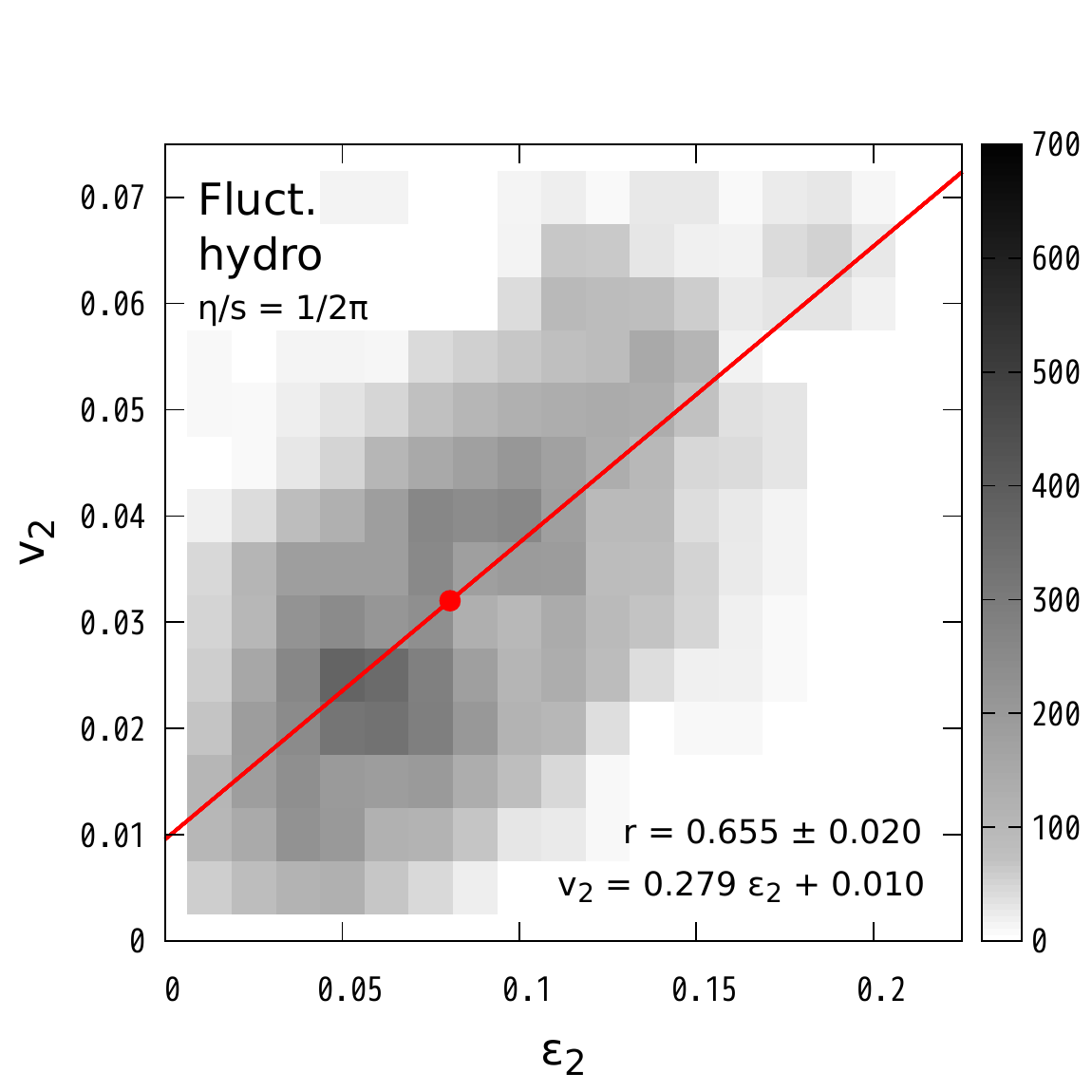}
    \includegraphics[clip,width=0.49\linewidth]{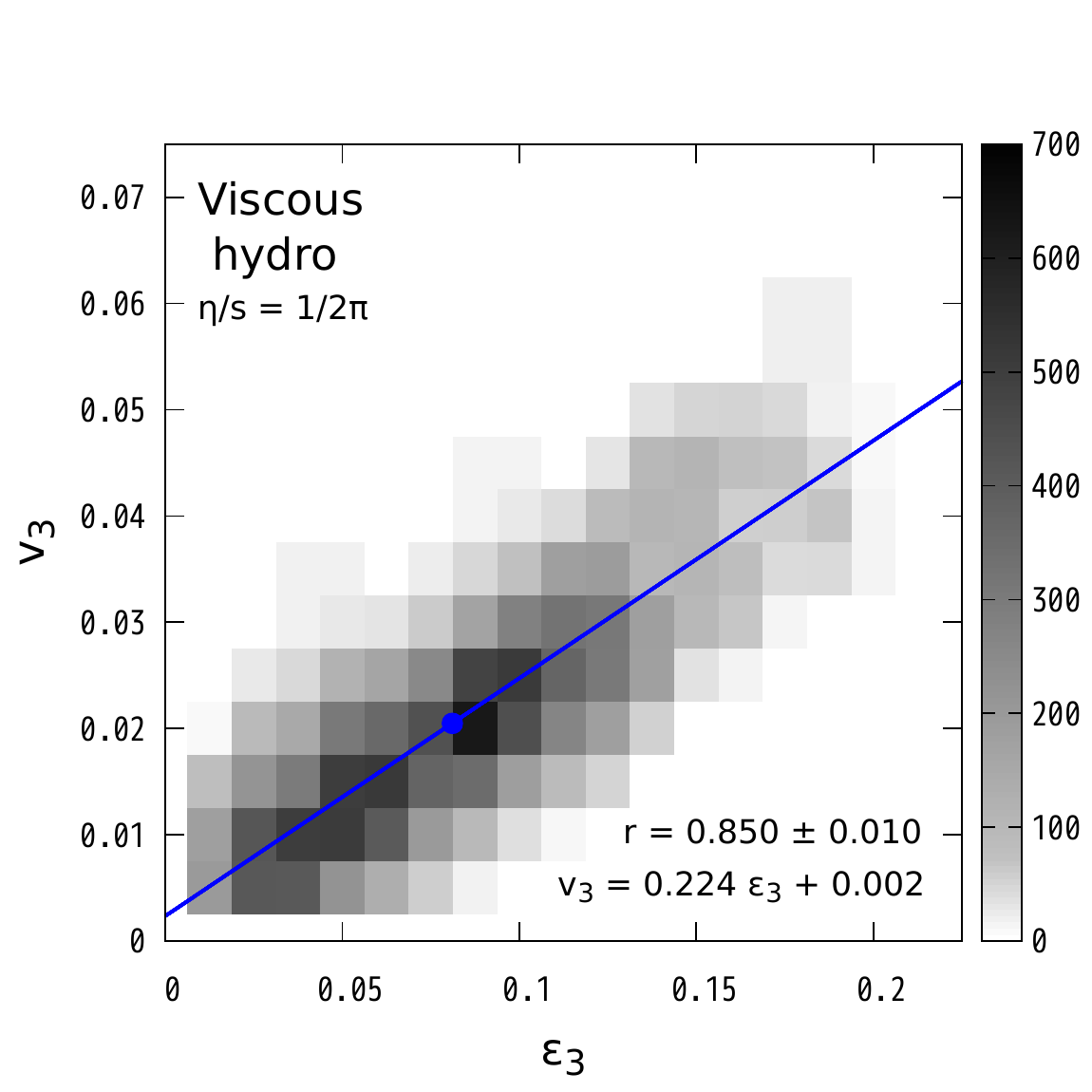}
    \includegraphics[clip,width=0.49\linewidth]{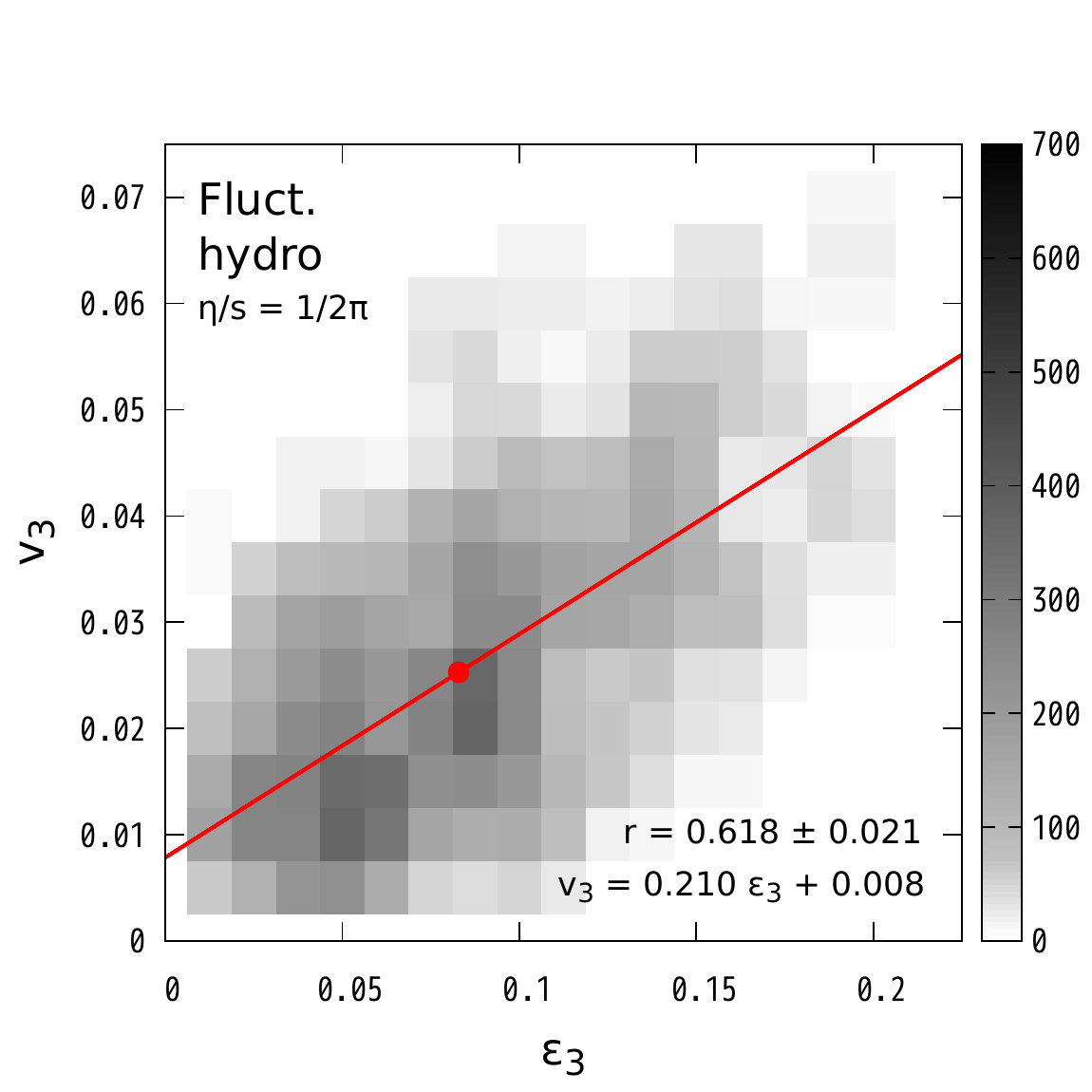}
    \caption{(Color Online)
            Correlation between event-by-event initial anisotropy $\varepsilon_n$ with $\eta_\text{s}=0$ and flow coefficient $v_n$ with $|\eta_p|<2.4$ in Pb+Pb collisions at $\sqrt{s_\text{NN}} = 2.76~\text{TeV}$ for 0--0.2\% centrality.
            Color bands correspond to the probability density.
            Results for $n=2$ (upper panels) and $3$ (lower panels) from viscous hydro (left panels) and fluctuating hydro (right panels) with specific shear viscosity $\eta/s=1/2\pi$ are shown.
            Linear fits (solid lines) and the Pearson correlation coefficients $r$ are also shown.
            }
    \label{fig:vn_scatter}
\end{figure}
Figure~\ref{fig:vn_scatter} shows the distribution of $\varepsilon_n$ and $v_n$ for $n=2$ and $3$ in Pb+Pb collisions at $\sqrt{s_\text{NN}} = 2.76~\text{TeV}$ for the 0--0.2\% centrality.
The event-by-event $v_n$ distributions are calculated by using Eq.~\eqref{eq:vn} without taking the event average but mixing the particles of 10 cascades from the same hydrodynamic event.
The geometrical anisotropies of the initial condition for a single event are calculated using the initial entropy density distribution~\eqref{eq:Glauber} as
\begin{equation}
    \label{eq:eccentricity}
    \varepsilon_n e^{in\Phi_n} = - \frac{\int r dr d\phi r^n e^{in\phi} s(r, \phi)}{\int r dr d\phi r^n s(r, \phi)},
\end{equation}
where $r$ and $\phi$ are the radius and the azimuthal angle, respectively, with the origin being the center-of-mass of the entropy density distribution, and $\Phi_n$ defines the $n$th-order participant-plane angle.
In Fig.~\ref{fig:vn_scatter}, we first notice that the linear relation is not perfect on an event-by-event basis but has fluctuations, $v_n = \kappa_n \varepsilon_n + \delta_n$.
This means that an initial-state analysis based on the linear relation between $v_n\{2\}$ and $\varepsilon_n\{2\} = \sqrt{\langle \varepsilon_n^2\rangle_\text{ev}}$ would suffer from the flow fluctuations as $v_n\{2\}^2 \sim \kappa_n^2 \varepsilon_n\{2\}^2 + \langle\delta_n^2\rangle_\text{ev}$, which calls for the necessity of the event-by-event dynamical calculations.
In Fig.~\ref{fig:vn_scatter}, we also see that hydrodynamic fluctuations increase the flow fluctuations of $v_n$ and result in fatter distribution with smaller values of the Pearson correlation coefficient $r$, which means that the linear relation $v_n = \kappa_n \varepsilon_n$ becomes even worse.
The hydrodynamic fluctuations also increase the intercept of the fitted line in Fig.~\ref{fig:vn_scatter}, i.e., the final anisotropy $v_n$ is generated by the hydrodynamic fluctuations even with the vanishing initial anisotropy $\varepsilon_n=0$.

These results suggest that the hydrodynamic fluctuations have a large impact on the linear relation $v_n=\kappa_n \varepsilon_n$, which are typically assumed in the flow-puzzle discussions based on the initial anisotropies, and thus we expect that the consideration of the hydrodynamic fluctuations would possibly change the situation.

\begin{figure}[tbp]
    \centering
    \includegraphics[clip,width=\linewidth]{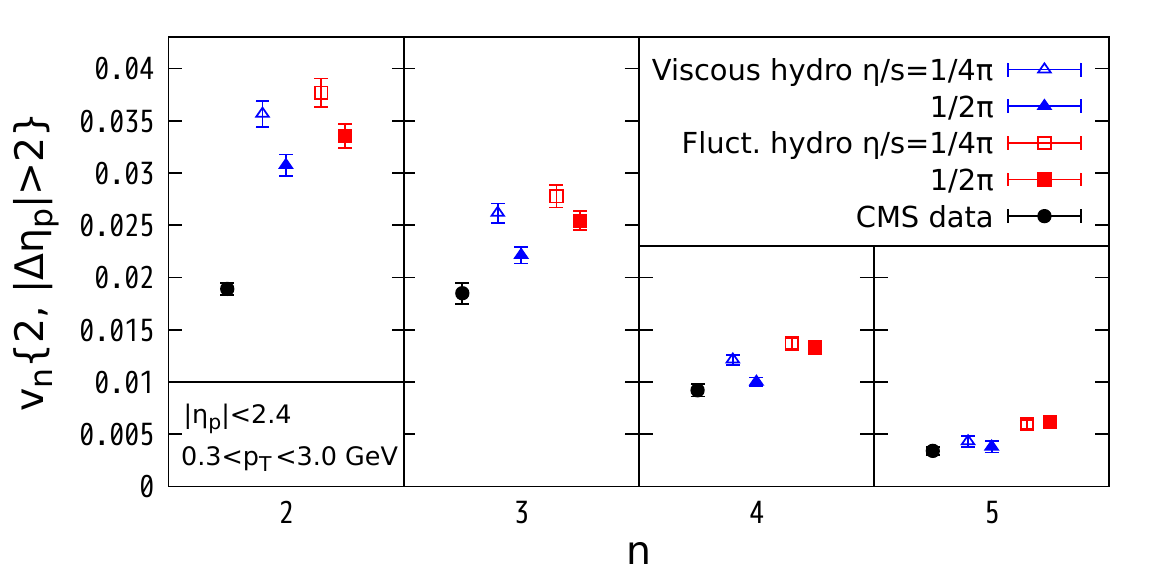}
    \caption{(Color Online)
            Charged-hadron anisotropic flow coefficients $v_n\{2\}$ up to fifth-order in Pb+Pb collisions at $\sqrt{s_{\rm NN}} = 2.76~\text{TeV}$ for 0--0.2\% centrality.
            The results from viscous hydrodynamics with $\eta/s=1/4\pi$ (open triangle) and $1/2\pi$ (filled triangle) and from fluctuating hydrodynamics with $\eta/s=1/4\pi$ (open square) and $1/2\pi$ (filled square) are compared with experimental data (filled circle) obtained by the CMS Collaboration~\cite{CMS:2013bza}.}
    \label{fig:vn}
\end{figure}
Figure~\ref{fig:vn} shows the anisotropic flow coefficients $v_n\{2\}$ with $|\Delta\eta_p|>2.0$ in Pb+Pb collisions at $\sqrt{s_{\rm NN}} = 2.76~\text{TeV}$ for 0--0.2\% centrality.
The overestimation of $v_n\{2\}$ seen in Fig.~\ref{fig:vn} is the known one in existing studies~\footnote{
  The overestimation might be suppressed by the bulk viscosity~\cite{Rose:2014fba}, the temperature-dependent shear viscosity larger at high $T$~\cite{Plumari:2015cfa}, and a different initial model that has smaller geometrical anisotropies than the MC-Glauber model~\cite{Shen:2015qta}.
  We do not address them in the present study.}.
In Fig.~\ref{fig:vn}, the hydrodynamic fluctuations are found to increase $v_n\{2\}$ due to the increased flow fluctuations of $v_n$ while the shear viscosity reduces the magnitudes of $v_n\{2\}$.
The increase by the hydrodynamic fluctuations is more significant for larger shear viscosity $\eta$ as naively expected from the FDR~\eqref{eq:FDR}.
We also observe in Fig~\ref{fig:vn} that the relative increase of $v_n$ by the hydrodynamic fluctuations is more significant in higher orders, which can be understood by the nature of the hydrodynamic fluctuations affecting smaller structures more.
This behavior qualitatively makes $v_3\{2\}$ closer to $v_2\{2\}$ and can improve the situation for the flow puzzle.
However, the effect is quantitatively too small to make $v_3\{2\}$ have the same magnitude as $v_2\{2\}$, and thus this does not quantitatively solve the flow puzzle.

\begin{figure}[tbp]
    \centering
    \includegraphics[clip,width=\linewidth]{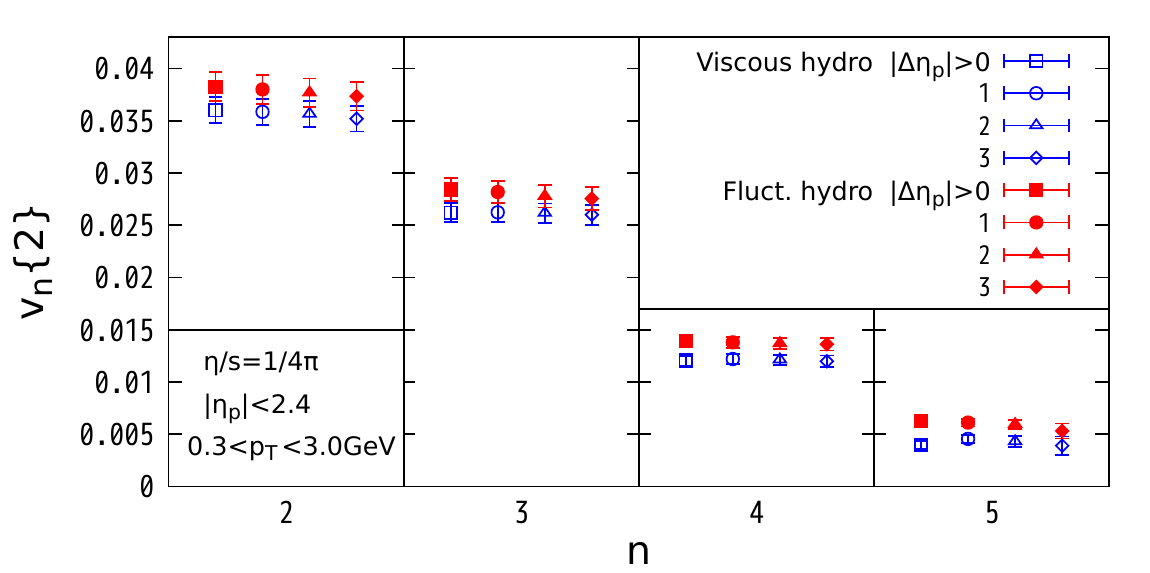}
    \includegraphics[clip,width=\linewidth]{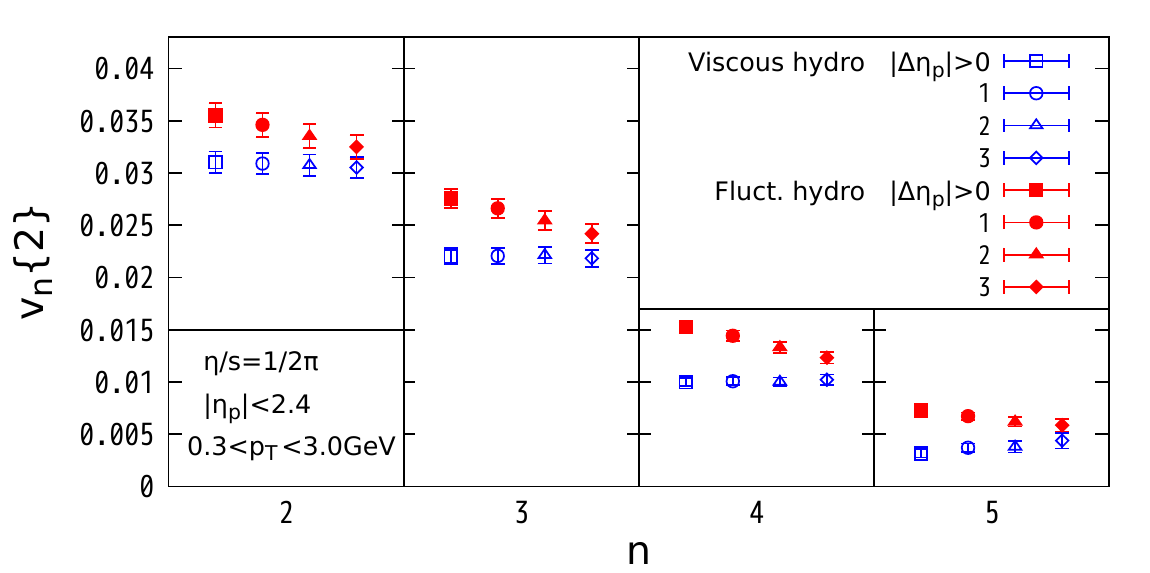}
    \caption{(Color Online)
            Pseudo-rapidity gap $|\Delta\eta_p|_{\min}$ dependence of $v_n\{2\}$ up to fifth-order in Pb+Pb collisions at $\sqrt{s_{\rm NN}} = 2.76~\text{TeV}$ for 0--0.2\% centrality.
            Results with $\eta/s=1/4\pi$ (top) and $1/2\pi$ (bottom) from viscous hydro (open symbols) and fluctuating hydro (filled symbols) for $|\Delta\eta_p|>0$ (square), $1$ (circle), $2$ (triangle), and $3$ (diamond) are shown.
            }
    \label{fig:vn_etagapdependence}
\end{figure}
To see the qualitative effects of hydrodynamic fluctuations in detail, we analyze the pseudo-rapidity gap $|\Delta\eta_p|_{\min}$ dependence of $v_n\{2\}$ in Fig.~\ref{fig:vn_etagapdependence}.
The results from the viscous hydro are nearly independent of $|\Delta\eta_p|_{\min}$ because the present model does not contain major physical sources of non-flow correlations such as jets.
On the other hand, in the case of fluctuating hydro, the influences of the hydrodynamic fluctuations on $v_n\{2\}$ become smaller as $|\Delta\eta_p|_{\min}$ increases, and $v_n\{2\}$ approaches that of the viscous hydro.
This is because the hydrodynamic fluctuations have larger effects in a short range, and taking large $|\Delta\eta_p|_{\min}$ reduces their effects.
Nevertheless, even with a large $|\Delta\eta_p|_{\min}$, there remain sizable effects of the hydrodynamic fluctuations, especially with large shear viscosity.
This means that taking the rapidity gap does not totally remove the effect of the hydrodynamic fluctuations.

\begin{figure}[tbp]
    \centering
    \includegraphics[clip,width=\linewidth]{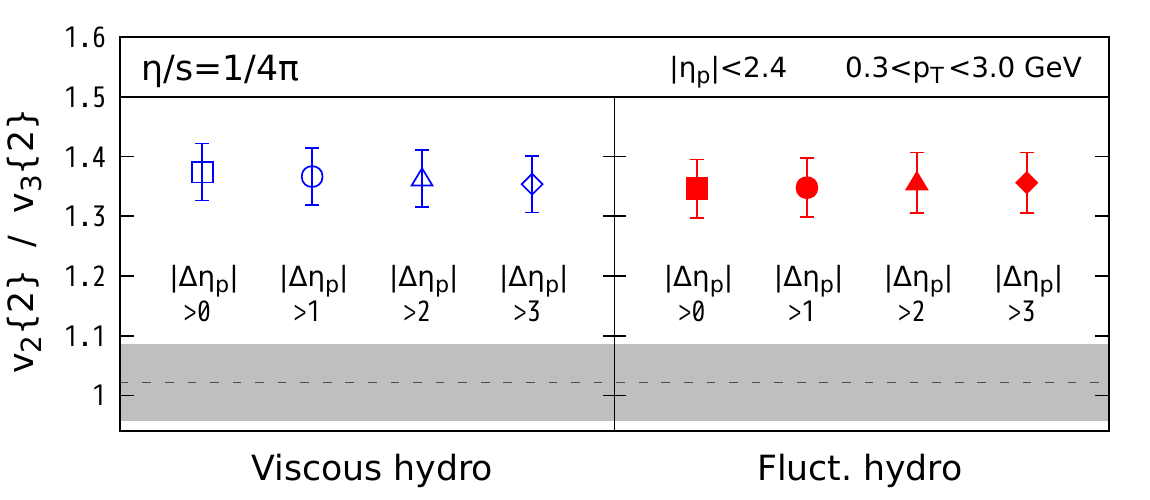}
    \includegraphics[clip,width=\linewidth]{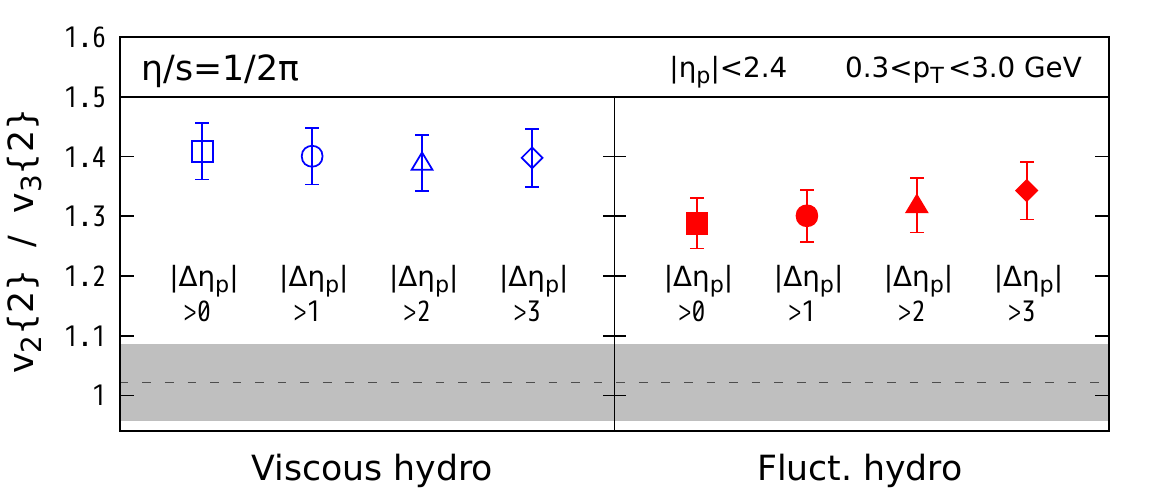}
    \caption{(Color Online)
            Pseudo-rapidity gap $|\Delta\eta_p|_{\min}$ dependence of the ratios $v_2\{2\}\,/\,v_3\{2\}$ with $\eta/s=1/4\pi$ (top) and $1/2\pi$ (bottom) in Pb+Pb collisions at $\sqrt{s_{\rm NN}} = 2.76~\text{TeV}$ for 0--0.2\% centrality.
            The symbols are the same as in Fig.~\ref{fig:vn_etagapdependence}.
            The experimental ratio $v_2\{2, |\Delta\eta_p|>2\}/v_3\{2, |\Delta\eta_p|>2\}\sim1.02$ (dashed line) with uncertainty (shaded band) calculated from Ref.~\cite{CMS:2013bza} is also shown\protect\footnotemark.}
    \label{fig:v2v3ratio}
\end{figure}
\footnotetext{The experimental uncertainty of the ratio is estimated by assuming uncorrelated uncertainties between $v_2\{2, |\Delta\eta_p|>2\}$ and $v_3\{2, |\Delta\eta_p|>2\}$ from Ref.~\cite{CMS:2013bza} because the experimental covariances are not available.
Therefore, the uncertainty is overestimated.}
Figure~\ref{fig:v2v3ratio} shows the $v_2\{2\}/v_3\{2\}$ ratios calculated from the results shown in Fig.~\ref{fig:vn_etagapdependence}.
With the shear viscosity $\eta/s= 1/4\pi$, the ratios with the viscous and fluctuating hydro have mostly the same value.
With the increased shear viscosity $\eta/s = 1/2\pi$, the ratio increases slightly with the viscous hydro but decreases slightly with the fluctuating hydro.
Also, the ratio approaches the experimental value with smaller $|\Delta\eta_p|_{\min}$, i.e., with larger effects of the hydrodynamic fluctuations.
The ratios of $|\Delta\eta_p|>2$ are also compared to the experimental value $v_2\{2, |\Delta\eta_p|>2\}/v_3\{2, |\Delta\eta_p|>2\}\sim1.02$ in Ref.~\cite{CMS:2013bza}. With the large shear viscosity $\eta/s = 1/2\pi$, the deviation of the ratio $\sim 1.39-1.02 = 0.37$ in the viscous hydro is reduced to $\sim 1.32-1.02 = 0.30$ in the fluctuating hydro by about 19\%.
Likewise, with the smaller shear viscosity $\eta/s = 1/4\pi$, the deviation is reduced by about 3\%, which is though not statistically significant.
This means that hydrodynamic fluctuations qualitatively contribute to resolving the ultra-central flow puzzle, especially with larger shear viscosity.
Nevertheless, it seems hard to quantitatively resolve the puzzle solely by the hydrodynamic fluctuations.

\vspace{10pt}
In this Letter, we investigated the effects of hydrodynamic fluctuations during the space-time evolution of the QGP fluid for the ultra-central flow puzzle using a (3+1)-dimensional integrated dynamical model with relativistic hydrodynamic model~\texttt{rfh} which includes causal hydrodynamic fluctuations and dissipation.
We used importance sampling to intensively generate ultra-central collision events. We performed simulations of 0--0.2\% central Pb+Pb collisions at $\sqrt{s_\mathrm{NN}} = 2.76~\text{TeV}$ with and without hydrodynamic fluctuations for comparison.
We showed that the hydrodynamic fluctuations worsen the linear relation $v_n = \kappa_n \varepsilon_n$ in the ultra-central collisions on an event-by-event basis so that the dynamical simulations are important for the quantitative analysis of the ultra-central flow puzzle.
We found that the hydrodynamic fluctuations increase the anisotropic flow coefficients $v_n\{2\}$ by increasing the flow fluctuations.
The hydrodynamic fluctuations also make the $v_2\{2\}/v_3\{2\}$ ratio closer to the experimental data, specifically by about 19\% with a larger shear viscosity $\eta/s = 1/2\pi$.
The effects of hydrodynamic fluctuations on $v_n\{2\}$ decrease with increasing pseudo-rapidity gap $|\Delta\eta_p|_{\min}$, yet there remain sizable effects of the hydrodynamic fluctuations even with a large rapidity gap $|\Delta\eta_p|_{\min}$.
Our analyses show that hydrodynamic fluctuations qualitatively improve the situation for the ultra-central flow puzzle, though their effects alone are too small to solve the puzzle within the present model.
In the future, we shall investigate the effect of hydrodynamic fluctuations with temperature-dependent shear viscosity, where we expect a larger effect coming from larger viscosity at high temperatures.
The ultra-central flow puzzle remains a challenge for hydrodynamic models even today, but the hydrodynamic fluctuations would certainly contribute to an improvement in the ultra-central flow ratio.

\vspace{10pt}
\section*{Acknowledgement}
The work by T.H. was partly supported by JSPS KAKENHI Grant No.~JP19K21881.

\bibliographystyle{elsarticle-num}
\biboptions{sort&compress}
\bibliography{reference}

\end{document}